\documentclass[12pt]{article}

\usepackage{amsmath,amssymb,cite,float,fullpage,graphicx,nicefrac,slashed,authblk}
\usepackage{multirow}
\usepackage[T1]{fontenc}
\usepackage[table]{xcolor}

\usepackage{color}
\definecolor{colorLink}{rgb}{0.9,0,0} 
\definecolor{colorCite}{rgb}{0,0.7,0} 
\definecolor{colorURL} {rgb}{0,0,0.8} 
\definecolor{colorCapt}{rgb}{0,0.7,0} 
\definecolor{colorPink}{rgb}{1,0,0.7} 
\usepackage[colorlinks=true,linktocpage=true,linkcolor=colorLink,citecolor=colorCite,urlcolor=colorURL]{hyperref}
\usepackage[labelfont={color=colorCapt,bf}]{caption}

\newcommand{\cL}          {\mathcal{L}}

\newcommand{\cO}          {\mathcal{O}}

\newcommand{\PW}{\mathrm{W}}
\newcommand{\sw}{s_{\scriptstyle\PW}}

\usepackage[normalem]{ulem}

\title{Non-Standard Neutrino Interactions at \mbox{Neutrino Experiments} and Colliders}
\author{Ayres Freitas}
\author{Matthew Low}
\affil{\small Pittsburgh Particle-physics Astro-physics \& Cosmology Center
(PITT-PACC),\\ Department of Physics \& Astronomy, University of Pittsburgh, Pittsburgh, PA 15260, USA}
\date{} 

\begin{document}

\maketitle

\begin{abstract}
The impact of new physics on the interactions of neutrinos with other particles can be parametrized by a set of effective four-fermion operators called non-standard neutrino interactions (NSIs).  This NSI framework is useful for studying the complementarity between different types of neutrino experiments.  In this work, we further compare the reach of neutrino experiments with high-energy collider experiments.  Since high-energy colliders often probe the mass scale associated with the four-fermion operators, the effective field theory approach becomes invalid and explicit models must be utilized.  We study a variety of representative simplified models including new U(1) gauge bosons, scalar leptoquarks, and heavy neutral leptons. For each of these, we examine the model parameter space constrained by NSI bounds from current and future neutrino experiments, and by data from the Large Hadron Collider and planned electron-positron and muon colliders. We find that in the models we study, with the possible exceptions of muon-philic leptoquarks and heavy neutral leptons mixing with electron neutrinos, collider searches are more constraining than neutrino measurements.  Additionally, we briefly comment on other model building possibilities for obtaining models where neutrino experiments are most constraining.
\end{abstract}


\section{Introduction}

Neutrino oscillation and scattering experiments can probe physics beyond the Standard Model (SM) through modifications of the effective interactions between neutrinos and matter particles (\emph{i.e.}\ nuclei and electrons)~\cite{Wolfenstein:1977ue}. These neutrino non-standard interactions (NSIs) can be parameterized in terms of a low-energy effective field theory framework with four-fermion operators (for reviews see Refs.~\cite{Ohlsson:2012kf,Farzan:2017xzy}),
\begin{align} \label{eq:NC}
\mathcal{L}_{\rm NC}
&= - 2 \sqrt{2} G_F 
\sum_{\alpha, \beta, f, P} \epsilon_{\alpha\beta}^{fP} (\bar{\nu}_\alpha \gamma^\mu P_L v_\beta)
(\bar{f} \gamma_\mu P f), \\
 \label{eq:CC}
\mathcal{L}_{\rm CC}
&= - 2 \sqrt{2} G_F 
\sum_{\alpha, \beta, f, P} \bar{\epsilon}_{\alpha\beta}^{fP} (\bar{\nu}_\alpha \gamma^\mu P_L \ell_\beta)
(\bar{f} \gamma_\mu P f') + \text{h.c.}
\end{align}
Here, $\alpha,\beta$ are generation indices, $f \in \{e,u,d\}$, and $P \in \{P_L,P_R\}$ with $P_{L/R} = \frac{1}{2}(1\mp \gamma_5)$.
Most neutrino experiments are more sensitive to NSIs between neutrinos and nuclei (rather than electrons). Therefore, this article will focus on $\epsilon_{\alpha\beta}^{fP}$ with $f=u,d$ (up-quarks/down-quarks).

Furthermore, we primarily consider neutral-current (NC) NSIs, Eq.~\eqref{eq:NC}.
It is convenient to define vector and axial-vector NSIs,
\begin{align}
\epsilon_{\alpha\beta}^u &\equiv \epsilon_{\alpha\beta}^{uL} + \epsilon_{\alpha\beta}^{uR},
&
\epsilon_{\alpha\beta}^d &\equiv \epsilon_{\alpha\beta}^{dL} + \epsilon_{\alpha\beta}^{dR}, \\
\epsilon_{\alpha\beta}^{Au} &\equiv \epsilon_{\alpha\beta}^{uR} - \epsilon_{\alpha\beta}^{uL},
&
\epsilon_{\alpha\beta}^{Ad} &\equiv \epsilon_{\alpha\beta}^{dR} - \epsilon_{\alpha\beta}^{dL}.
\end{align}
There are three main types of processes at neutrino experiments that are sensitive to neutrino NSIs: \emph{(i)} Long-baseline neutrino oscillations, where Eq.~\eqref{eq:NC} modifies the effective neutrino propagator; \emph{(ii)} coherent elastic neutrino-nucleus scattering (CE$\nu$NS) in a neutrino detector; and \emph{(iii)} inelastic neutrino-nucleus scattering, including deep inelastic scattering (DIS).
Neutrino oscillations and CE$\nu$NS are only sensitive to the vector-like interactions $\epsilon_{\alpha\beta}^q$, whereas axial-vector interactions $\epsilon_{\alpha\beta}^{Aq}$ can only be probed in inelastic scattering processes.

If the NSIs originate from beyond-the-SM (BSM) physics above the weak scale, they can be described in terms of higher-dimensional operators in an effective field theory extension of the SM (SMEFT). The leading contribution to NC NSIs comes from dimension-6 operators of the form~\cite{Gavela:2008ra} 
\begin{align}
\cO_{Lq}^{(1)} &= (\overline{L}_\alpha \gamma^\mu L_\beta) (\bar{q} \gamma_\mu q), \label{eq:Olq} \\
\cO_{Lq}^{(3)} &= (\overline{L}_\alpha \gamma^\mu \tau^a L_\beta) (\overline{Q} \gamma_\mu \tau_a Q), \label{eq:Olq3} \\
\cO_{HL}^{(1)} &= (H^\dagger i \!\stackrel{\leftrightarrow}{D}_\mu\! H) (\overline{L}_\alpha \gamma^\mu L_\beta) \label{eq:OHl}, \\
\cO_{HL}^{(3)} &= (H^\dagger i \!\stackrel{\leftrightarrow}{D}_\mu\! \tau_a H) (\overline{L}_\alpha \gamma^\mu \tau^a L_\beta) \label{eq:OHl3},
\end{align}
where $\tau_a$ are the SU(2) generators. Here $q$ refers to either a right-handed up or down quark or to the quark doublet, while $Q$ denotes only a quark doublet. Upon integrating out weak-scale degrees of freedom, these SMEFT operators translate into the low-energy Lagrangians of Eq.~\eqref{eq:NC} and Eq.~\eqref{eq:CC}.

Any non-diagonal ($\alpha \neq \beta$) flavor components of these SMEFT operators are very strongly constrained by charged lepton flavor violating processes, and thus they are negligible for neutrino physics. Therefore, only flavor diagonal ($\alpha=\beta$) BSM interactions will be considered in the following.

Even in the flavor-conserving case, the operators Eqs.~\eqref{eq:Olq}--\eqref{eq:Olq3} necessarily produce deviations in processes with charged leptons that can be probed at colliders, \emph{e.g.}\ $q\bar{q}\to\ell^+\ell^-$ or $\ell^+\ell^- \to q\bar{q}$. However, at high-energy colliders, such as the Large Hadron Collider (LHC), the effective theory description in the SMEFT framework is not necessarily valid, and one instead needs to explicitly take into account the fields that mediate these operators. In other words, bounds on NSIs from colliders are model-dependent.

To avoid constraints from processes with charged leptons, one could consider the possibility that the neutrino NSIs are produced by dimension-8 SMEFT operators~\cite{Berezhiani:2001rs,Gavela:2008ra,Babu:2020nna}, such as
\begin{align}
    \cO_{HLq}^{(1)} &= (\overline{L}_\alpha H) \gamma^\mu (H^\dagger L_\beta) (\bar{q} \gamma_\mu q). \label{eq:OHlq}
\end{align}
Once the Higgs boson acquires a vacuum expectation value (vev), Eq.~\eqref{eq:OHlq} will lead to 4-fermion interactions for neutrinos, but not for charged leptons at tree-level. Unfortunately, it is difficult to find UV-complete models that generate an operator of the form in Eq.~\eqref{eq:OHlq} without any of the dimension-6 operators Eqs.~\eqref{eq:Olq}--\eqref{eq:OHl3}. It was shown in Ref.~\cite{Gavela:2008ra} that for tree-level mediators, one needs a fine-tuned cancellation between contributions from two different mediator fields to suppress the dimension-6 operators. As explained below, we find that a similar conclusion holds if these operators are generated at one-loop level. In addition, we investigate to what extent the collider bounds, \emph{e.g.}\ from Drell-Yan production ($q\bar{q} \to \ell^+\ell^-$), are weakened in such a fine-tuned UV completion.

The purpose of the present paper is to analyze the relationship between neutrino NSIs and collider phenomenology for a representative set of minimal models, \emph{i.e.}\ models that introduce one BSM mediator at a time for NSIs. There are already many examples of such complementarity studies in the literature~\cite{Berezhiani:2001rs, Barranco:2007tz, Honda:2007wv,Antusch:2008tz,BuarqueFranzosi:2015qil, Choudhury:2018xsm, Pandey:2019apj, Babu:2019mfe, Han:2019zkz, Babu:2020nna, Han:2020pff, Cheung:2021tmx, Jana:2023ogd}. This paper goes beyond these studies by \emph{(a)} considering a broader range of models with spin-0, spin-1/2, and spin-1 mediators, \emph{(b)} taking into account current and future bounds from neutrino oscillations and neutrino scattering, and \emph{(c)} considering constraints from hadron colliders (LHC), future $e^+e^-$ colliders (such as FCC-ee), and future multi-TeV muon colliders. For concreteness, we only consider models with mediator masses at or above the weak scale (see \emph{e.g.}\ Refs.~\cite{Harnik:2012ni,Farzan:2015hkd,Farzan:2016wym,Babu:2017olk,Han:2019zkz,Li:2022jfl,Dutta:2024hqq} for other studies focused on low-mass mediators). For the collider phenomenology, as mentioned above, it is important to study explicit models rather than using an effective operator framework.

After reviewing the most relevant constraints on NSIs from neutrino experiments in section~\ref{sec:neuexp} and the complementary observables from colliders in section~\ref{sec:coll}, we analyze the current and projected future bounds on three classes of typical NSI mediators: U(1) gauge bosons (sections~\ref{sec:ZBL}, \ref{sec:Zax}), leptoquarks (section~\ref{sec:leptoquark}), and heavy neutral leptons (section~\ref{sec:HNL}). In section~\ref{sec:dim8} we discuss the scenario in which NSIs are generated by dimension-8 operators, using a concrete model example to study the impact on the collider bounds. We present our summary in section~\ref{sec:summ}.


\section{Neutrino experiments}
\label{sec:neuexp}


\subsection{Current neutrino data}

Neutrino NSIs impact the propagation of neutrinos through matter, and thus have an effect on neutrino oscillation phenomena, for example in long-baseline experiments like MINOS, T2K, and NO$\nu$A. Some recent analyses of NSIs from combinations of oscillation data from a range of experiments can be found in Refs.~\cite{Farzan:2017xzy,Du:2020dwr}.
However, neutrino oscillations by themselves are only sensitive to lepton flavor \emph{differences} among NSI parameters, \emph{e.g.} $\epsilon^q_{ee}-\epsilon^q_{\mu\mu}$ and $\epsilon^q_{\mu\mu}-\epsilon^q_{\tau\tau}$. This degeneracy can be broken by including neutrino scattering data, where currently the most constraining information stems from the measurement of CE$\nu$NS observed by the COHERENT experiment.

In this work, we utilize the combination of solar, atmospheric, reactor, short baseline, and long baseline neutrino experiments with COHERENT from Ref.~\cite{Coloma:2023ixt},\footnote{See Refs.~\cite{Esteban:2018ppq,Coloma:2019mbs,Dutta:2020che,Gehrlein:2024vwz} for examples of other combined analyses of neutrino oscillation and scattering data.} which leads to the constraints summarized in Tab.~\ref{tab:limits1}. Besides the allowed ranges listed in the table, the global analysis of Ref.~\cite{Coloma:2023ixt} identifies a second set of allowed values for $\epsilon^q_{\ell\ell}$ of ${\cal O}(0.3 - 0.4)$. However, such large values of the NSI parameters are excluded by other constraints for all the models considered in this paper (including neutrino scattering data from NO$\nu$A~\cite{Gehrlein:2024vwz}), which is why we will not mention these alternative solutions in the following.

For muon neutrinos, more stringent constraints on the NSI parameters can be obtained from NuTeV scattering data~\cite{Davidson:2003ha}, which are also listed in Tab.~\ref{tab:limits1}. Note that, for the sake of uniformity with bounds from other experiments, the 90\% confidence level (C.L.) results from Refs.~\cite{Coloma:2023ixt,Davidson:2003ha} have been translated to 95\% C.L. assuming Gaussian error distributions.\footnote{Comparing to the posterior distributions of Ref.~\cite{Coloma:2023ixt}, our translation to 95\% C.L. results in an underestimation of the size of the conference interval.  Therefore, the actual 95\% C.L. bounds from oscillation and CE$\nu$NS data are weaker than we show.}  The CHARM experiment~\cite{CHARM:1987pwr} and  the CDHS experiment~\cite{Blondel:1989ev} have also set bounds on muon neutrino scattering on nuclei, but they are weaker than those from NuTeV~\cite{Escrihuela:2011cf}.

\begin{table}
\centering
\begin{tabular}{cc}
    \hline
    Parameter & Allowed range \\
    \hline
    $\epsilon^u_{ee}$ & $[-0.045,0.041]$ \\ 
    $\epsilon^u_{\mu\mu}$ & $[-0.055,0.037]$ \\ 
    $\epsilon^u_{\tau\tau}$ & $[-0.055,0.039]$ \\ 
    $\epsilon^d_{ee}$ & $[-0.043,0.037]$ \\ 
    $\epsilon^d_{\mu\mu}$ & $[-0.048,0.045]$ \\ 
    $\epsilon^d_{\tau\tau}$ & $[-0.049,0.051]$ \\ 
    \hline
\end{tabular}
\hspace{5em}
\begin{tabular}{cc}
    \hline
    Parameter & Allowed range \\
    \hline
    $\epsilon^{uL}_{\mu\mu}$ & $[-0.004,0.004]$ \\ 
    $\epsilon^{uR}_{\mu\mu}$ & $[-0.010,0.004]$ \\ 
    $\epsilon^{dL}_{\mu\mu}$ & $[-0.004,0.004]$ \\ 
    $\epsilon^{dR}_{\mu\mu}$ & $[-0.010,0.018]$ \\ 
    \hline
\end{tabular}
\caption{Limits on NSI parameters from current neutrino experiments at 95\% C.L., assuming Gaussian error distributions. 
\emph{Left:} Combination of neutrino oscillation and CE$\nu$NS data from Tab.~3 of Ref.~\cite{Coloma:2023ixt}. \emph{Right:} Bounds from NuTeV neutrino scattering data~\cite{Davidson:2003ha}.}
\label{tab:limits1}
\end{table}

\subsection{Future neutrino data}
\label{sec:nufut}

Future neutrino oscillation data from DUNE, T2HK, and JUNO are not expected to lead to improved bounds on the NSI parameters~\cite{Liao:2016orc,Chatterjee:2021wac,Du:2021rdg,Cherchiglia:2023aqp}. However, neutrino scattering in the DUNE near detector (ND) can produce tighter limits. We use the work of Ref.~\cite{Falkowski:2018dmy} and Ref.~\cite{Abbaslu:2023vqk} for projected bounds on vector and axial-vector NSIs, respectively.

The limits in Ref.~\cite{Falkowski:2018dmy} are presented in terms of effective couplings $g^{\nu_\mu}_L$ and $g^{\nu_\mu}_R$ (see Fig.~2 therein) that combine up-quark and down-quark contributions and are normalized to the CC neutrino scattering rate. Assuming no NSIs in the CC neutrino interactions, these effective couplings relate to the NC NSI parameters as follows:
\begin{align}
    |\epsilon^u_{\mu\mu}| &= (1+h)\, \biggl|\frac{g^\nu_{\rm L,SM}}{g^{\nu u}_{\rm LL,SM} + r_L g^{\nu d}_{\rm LL,SM}} \biggr| \, |\delta g^{\nu_\mu}_L|, \label{eq:epsgu} \\
    |\epsilon^d_{\mu\mu}| &= (r_L + h r_R)\, \biggl| \frac{g^\nu_{\rm L,SM}}{g^{\nu u}_{\rm LL,SM} + r_L g^{\nu d}_{\rm LL,SM}} \biggr| \, |\delta g^{\nu_\mu}_L|, \label{eq:epsgd}
\end{align}
where 
\begin{align}
    &g^{\nu u}_{\rm LL,SM} = \tfrac{1}{2}-\tfrac{2}{3}\sw^2 \approx 0.34, \qquad g^{\nu d}_{\rm LL,SM} = -\tfrac{1}{2}+\tfrac{1}{3}\sw^2 \approx -0.42, \\ 
    &g^\nu_{\rm L,SM} = \sqrt{(g^{\nu u}_{\rm LL,SM})^2+(g^{\nu d}_{\rm LL,SM})^2} \approx 0.54,
\end{align}
while $r_P \equiv \epsilon^{dP}_{\mu\mu}/\epsilon^{uP}_{\mu\mu}$ ($P=L,R)$ and $h \equiv \epsilon^{uR}_{\mu\mu}/\epsilon^{uL}_{\mu\mu}$ encapsulate the model dependence.  The parameter $s_W$ is the sine of the Weinberg angle.

The limit on $\delta g^{\nu_\mu}_L$ is extracted from Fig.~2 of Ref.~\cite{Falkowski:2018dmy} by examining the contours along a line $\frac{\delta g^{\nu_\mu}_R}{\delta g^{\nu_\mu}_L} = \frac{g^\nu_{\rm R,SM}}{g^\nu_{\rm L,SM}} \times h$, with $g^\nu_{\rm R,SM} = \sqrt{(-\tfrac{2}{3}\sw^2)^2+(\frac{1}{3}\sw^2)^2} \approx 0.18$. We will consider projections from Ref.~\cite{Falkowski:2018dmy} based \emph{(i)} only on statistical uncertainties, and \emph{(ii)} including a 0.1\% systematic uncertainty on the measurement of the NC to CC scattering rate ratio.

Ref.~\cite{Abbaslu:2023vqk} presents projected limits for the DUNE ND directly on the NSI parameters $\epsilon^{Aq}_{\mu\mu}$ in Tab.~VI. Again, we consider two scenarios considered in that work: \emph{(i)} only statistical uncertainties, and \emph{(ii)} including a 10\% systematic uncertainty on the detection efficiency for NC scattering events. The projected limits, translated to 95\% C.L., are summarized in Tab.~\ref{tab:limits2}.

\begin{table}
\centering
\begin{tabular}{ccc}
    \hline
    Parameter & Stat.\ uncertainty & Stat.+Syst.\ uncertainty \\
    \hline
    $\epsilon^{Au}_{\mu\mu}$ & $\pm 1.2\times 10^{-4}$ & $\pm 0.8 \times 10^{-3}$ \\
    $\epsilon^{Ad}_{\mu\mu}$ & $\pm 1.1 \times 10^{-4}$ & $\pm 4.1\times 10^{-3}$\\
    \hline
\end{tabular}
\caption{Projected bounds on axial-vector NSI parameters from neutrino-nucleus scattering in the DUNE ND~\cite{Abbaslu:2023vqk} at 95\% C.L. For the systematic uncertainty a 10\% error on the detection efficiency of NC scattering events has been assumed.}
\label{tab:limits2}
\end{table}

The proposed PINGU upgrade~\cite{IceCube:2016xxt} to the IceCube experiment and the ORCA detector of the KM3NeT experiment \cite{KM3Net:2016zxf} have sensitivity to $\mu$- and $\tau$-flavor neutrinos, which for the $\tau$-flavor cases exceed the reach of accelerator-based experiments.  We use the projected bounds of Ref.~\cite{HernandezRey:2021qac} on $\epsilon^\oplus_{\tau\tau}$, the NSI for Earth matter, for ORCA (which are stronger than estimated bounds for PINGU \cite{Choubey:2014iia}), for BSM scenarios that specifically couple to $\tau$ neutrinos. The Earth matter NSIs can be converted into quark-level NSIs through the relation~\cite{Coloma:2023ixt}
\begin{equation}
\label{eq:conv-earth-matter}
\epsilon^\oplus_{\alpha\beta} = 
(2 + Y_n^\oplus) \epsilon^u_{\alpha\beta}
+ (1 + 2 Y_n^\oplus) \epsilon^d_{\alpha\beta},
\end{equation}
where $Y_n^\oplus$ is the neutron-proton ratio that has an average value of $Y_n^\oplus = 1.051$ in the PREM model~\cite{Dziewonski:1981xy}. The bounds on $\epsilon^\oplus_{\tau\tau}$ in Ref.~\cite{HernandezRey:2021qac} depend on the neutrino hierarchy; we choose the weaker of the bounds to be conservative, resulting in $|\epsilon^\oplus_{\tau\tau}| < 0.0134$ at 95\% C.L.\ assuming Gaussian error distribution.

Note that all of the systematic errors quoted above are only for illustration purposes and not based on detailed experimental studies.


\section{Constraints from colliders}
\label{sec:coll}

Any model that generates any of the operators in Eqs.~\eqref{eq:Olq}-\eqref{eq:OHl3} at low energies will not only produce neutrino NSIs,
but necessarily also lead to BSM contributions to the processes $q\bar{q} \to \ell^+\ell^-$ and $\ell^+\ell^- \to q\bar{q}$, where $\ell$ is a charged lepton. 
Depending on the model, lepton scattering processes, $\ell^+\ell^- \to \ell^{\prime +}\ell^{\prime -}$, can also be affected.
As a result, these models can be constrained by measurements of Drell-Yan production at hadron colliders, such as the LHC, and fermion pair production processes at lepton colliders, such as LEP.

For the LHC, we will map constraints from recent ATLAS and CMS studies to the parameter space of the models considered in this paper, and also use projections for the HL-LHC from the experimental collaborations. In addition, there are model-dependent limits from direct searches for BSM particles at the LHC, which will be discussed in the results sections below.

For lepton colliders, bounds from fermion pair production at LEP2 turn out not to be competitive due to the limited energy reach and statistics. However, future $e^+e^-$ colliders such as ILC~\cite{ILC:2013jhg,Bambade:2019fyw}, FCC-ee~\cite{FCC:2018evy}, and CEPC~\cite{CEPCStudyGroup:2018ghi,CEPCStudyGroup:2023quu} have much improved sensitivity due to their high luminosity. Moreover, a high-energy lepton collider, in particular a multi-TeV muon collider~\cite{Accettura:2023ked}, would have unique capabilities to test models that generate NSIs.

To estimate the prospective reach of future $e^+e^-$ and $\mu^+\mu^-$ colliders, we compute cross-sections and forward-backward asymmetries using two different software frameworks for independent cross-checks: {\tt CalcHEP 3.4.6}~\cite{Belyaev:2012qa} and {\tt MadGraph 5}~\cite{Alwall:2014hca}. The calculations with {\tt CalcHEP} include initial-state QED radiation using the structure function approach of Kuraev and Fadin~\cite{Kuraev:1985hb}.

\medskip\noindent
For studies of $e^+e^- \to \mu^+\mu^-$ and $e^+e^- \to jj$ (where $j$ denotes a light-flavor quark-induced jet) we assume:
\begin{itemize}
\item Center-of-mass energy $\sqrt{s}=240$~GeV;
\item Total integrated luminosity $\cL_\text{int}=5 \mathrm{\,ab}^{-1}$ (which corresponds to the FCC-ee run plan~\cite{Bernardi:2022hny});
\item Muon efficiency: 99\%;
\item Jet efficiency: 99\%;
\item Absolute luminosity uncertainty: $5\times 10^{-4}$;
\item Generic cuts, based on Ref.~\cite{ALEPH:2013dgf} section 3.2: 
\newline
$m_{ff}>0.85\sqrt{s}, \quad |\cos\theta_f| <0.95$\quad (where $f=\mu,j$, $m_{ff}$ is the di-muon or di-jet invariant mass, and $\theta_f$ is the angle of $f$ relative to the beam axis).

\end{itemize}
On the other hand, the following parameters are used for studies of $\mu^+\mu^- \to e^+e^-$ and $\mu^+\mu^- \to jj$:
\begin{itemize}
\item Center-of-mass energy $\sqrt{s}=3$~TeV;
\item Total integrated luminosity $\cL_\text{int}=1 \mathrm{\,ab}^{-1}$ \cite{Narain:2022qud};
\item Electron efficiency: 90\% (see Figs.~77, 78 in Ref.~\cite{Accettura:2023ked});
\item Jet efficiency: 95\% (see Fig.~52 in Ref.~\cite{Accettura:2023ked});
\item Absolute luminosity uncertainty: $2\times 10^{-3}$ (see Sec.~4.5 in Ref.~\cite{Accettura:2023ked});
\item Generic cuts (see \emph{e.g.}\ Ref.~\cite{Accettura:2023ked} Figs.~53, 78): 
\newline
$
m_{ff}>0.85\sqrt{s}, \quad |\cos\theta_f|<0.87 \quad (f=e,j).
$
\end{itemize}
We only include irreducible SM contributions, since other backgrounds are expected to be small (the invariant mass cut reduces background from two-photon collisions).

Besides total cross-sections (for all final-states) we also consider the forward-backward asymmetry for $e^+e^-$ or $\mu^+\mu^-$ final states.
We do not consider the processes $e^+e^- \to e^+e^-$ or $\mu^+\mu^- \to \mu^+\mu^-$, which have larger SM backgrounds, and $\tau^+\tau^-$ final states, which have lower reconstruction efficiencies.


\section{Results for minimal models}
\label{sec:models}

\subsection{\boldmath $B{-}L$ gauge boson}
\label{sec:ZBL}

For illustrative purposes, let us begin with an extension of the SM with a $Z'_{B-L}$ gauge boson. The interplay of neutrino and collider phenomenology for models of this kind has been studied in the literature before (see \emph{e.g.}\ Refs.~\cite{Han:2019zkz,Babu:2020nna}), and we do not have any qualitatively new insights, but the structural simplicity of the model makes it useful to describe how we derive bounds on the parameter space.

The U(1)$_{B-L}$ gauge interactions of the quarks and leptons are given by
\begin{align}
    \cL_{B-L}^{\rm int} &= -g_{Z'}\sum_q \tfrac{1}{3} \bar{q}\gamma^\mu q \, Z'_\mu -g_{Z'}\sum_\ell (-1) \bar{\ell}\gamma^\mu \ell \, Z'_\mu\,,
\end{align}
where $Z'$ is the U(1)$_{B-L}$ gauge boson with gauge coupling $g_{Z'}$ and mass $m_{Z'}$.  Matching to the NSIs parameters yields
\begin{align}
\epsilon^{q}_{\ell\ell} &= -\frac{g_{Z'}^2}{3\sqrt{2} G_F m_{Z'}^2},
&
\epsilon^{Aq}_{\ell\ell} &= 0,
&
&(q = u,d; \; \ell=e,\mu,\tau).
\end{align}
As $\epsilon^{q}_{\ell\ell}$ are always negative in the $Z'_{B-L}$ model, the lower bounds in Tab.~\ref{tab:limits1} apply. Furthermore, given the universal couplings to up-type and down-type quarks in this model, we can combine the bounds on $\epsilon^u_{\ell\ell}$ and $\epsilon^d_{\ell\ell}$ to arrive at the stronger limits (95\% C.L.)
\begin{align}
  \epsilon^q_{ee} &> -0.031 \qquad \text{($\nu$ osc. + CE$\nu$NS)}, \\
  \epsilon^q_{\mu\mu} &> -0.036 \qquad \text{($\nu$ osc. + CE$\nu$NS)}, \\
  \epsilon^q_{\mu\mu} &> -0.007 \qquad \text{(NuTeV)},
\end{align}
where for the NuTeV we used the fact that $\epsilon^{qL}_{\ell\ell} = \frac{1}{2}\epsilon^{q}_{\ell\ell}$ in the $Z'_{B-L}$ model.

For the projected reach of DUNE ND scattering, we derive bounds from Ref.~\cite{Falkowski:2018dmy} using the procedure outlined in section~\ref{sec:nufut}. One has $r_L=h=1$ in the $Z'_{B-L}$ model, so that one has to consider a line $\frac{\delta g^{\nu_\mu}_R}{\delta g^{\nu_\mu}_L} \approx 0.33$ in Fig.~2 of Ref.~\cite{Falkowski:2018dmy}, from which we obtain
\begin{align}
  |g_L^{\nu_\mu}| &< 0.0001 \qquad \text{(stat. error only)}, \\
  |g_L^{\nu_\mu}| &< 0.0005 \qquad \text{(0.1\% syst. error)}.
\end{align}
These bounds can be translated to bounds on the NSI parameters using Eqs.~\eqref{eq:epsgu} and ~\eqref{eq:epsgd}, yielding
\begin{align}
        |\epsilon^q_{\mu\mu}| &= 2\biggl|\frac{g_{\rm L,SM}^\nu}{g^{\nu u}_{\rm LL,SM} + g^{\nu d}_{\rm LL,SM}}\biggr|\, |\delta g_L^\nu| \approx 13.6\, |\delta g_L^\nu|.
\end{align}
At the LHC, this model is most severely constrained through direct searches for new gauge bosons in Drell-Yan production, $pp \to Z' \to \ell^+\ell^-$ ($\ell=e,\mu$). We use limits from ATLAS~\cite{ATLAS:2019erb,ATLAS:2019erb_repo} and CMS~\cite{CMS:2021ctt,CMS:2021ctt_repo} for a center-of-mass energy of 13~TeV and a luminosity of 139--140~fb$^{-1}$. Those limits are given in terms of a sequential $Z'$, which we recast to the $Z'_{B-L}$ model by rescaling the cross-sections calculated with  {\tt CalcHEP 3.4.6} \cite{Belyaev:2012qa} and {\tt MadGraph 5} \cite{Alwall:2014hca}, using a model file from Ref.~\cite{Amrith:2018yfb,Deppisch:2018eth,Basso:2008iv}\footnote{The model files used are from \url{https://feynrules.irmp.ucl.ac.be/wiki/B-L-SM}.} for {\tt FeynRules}~\cite{Alloul:2013bka} and {\tt CTEQ6L} PDFs~\cite{Pumplin:2002vw}.

Similarly, projections for the HL-LHC with 14~TeV center-of-mass energy and 3~ab$^{-1}$ luminosity can be obtained from Refs.~\cite{ATLAS:2018tvr,CMS:2022gho}.

\medskip
The main constraints from future $e^+e^-$ (muon) colliders stem from $e^+e^- \to f\bar{f}$ ($\mu^+\mu^- \to f\bar{f}$) processes via off-shell $Z'_{B-L}$ exchange. We analyze these processes as described in section~\ref{sec:coll}.

\begin{figure}[t]
    \centering
    \includegraphics[width=0.75\linewidth]{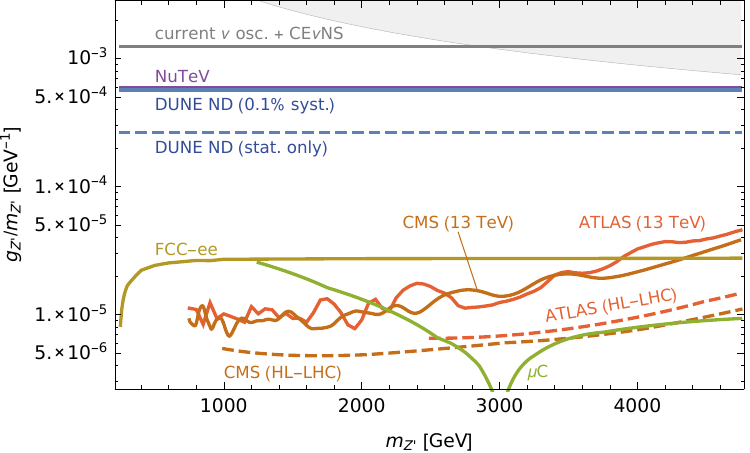}
    \caption{Comparison of current and future bounds (95\% C.L.) from neutrino experiments and colliders on $g_{Z'}/m_{Z'}$ for the $Z'_{B-L}$ model. See text for more details. The shaded region displays the perturbativity bound.}
    \label{fig:ZBL}
\end{figure}

Figure~\ref{fig:ZBL} shows the resulting comparison of current and future constraints from NSIs at neutrino experiments and $Z'_{B-L}$ signatures at colliders. Existing LHC constraints are already significantly stronger than the projected sensitivity of neutrino scattering at DUNE, even in the scenario without systematic uncertainties. The HL-LHC and a future muon collider can further improve these constraints. The gray shaded region in the plot indicates the region where the U(1)$_{B-L}$ gauge coupling becomes non-perturbative, $g_{Z'} \gtrsim \sqrt{4\pi}$.


\subsection{Gauge boson with axial charges}
\label{sec:Zax}

The U(1)$_{B-L}$ model produces only vector NSIs. For comparison, it is interesting to consider a $Z'$ with significant axial-vector couplings to quarks. We use the framework for anomaly-free U(1) extensions of the SM from Ref.~\cite{Appelquist:2002mw}, with an 
interaction Lagrangian of the form
\begin{align}
    \cL_{\rm ax}^{\rm int} &= -g_{Z'}\sum_f z_f\, \bar{f}\gamma^\mu f \, Z'_\mu\,,
\end{align}
where $f=Q,L$ for the left-handed fermions and $f=u,d,e,\nu$ for the right-handed fermions. We choose the charges
\begin{equation}
z_Q = -\tfrac{1}{3},
\qquad
z_u = \tfrac{2}{3},
\qquad
z_d = -\tfrac{4}{3},
\qquad
z_L = 1,
\qquad
z_e = 0,
\qquad 
z_\nu = 2.
\end{equation}
Matching to the NSI parameters leads to
\begin{align}
\epsilon^{u}_{\ell\ell} &= \frac{g_{Z'}^2}{6\sqrt{2} G_F m_{Z'}^2},
&
\epsilon^{d}_{\ell\ell} &= -\frac{5g_{Z'}^2}{6\sqrt{2} G_F m_{Z'}^2},
&
\epsilon^{Au}_{\ell\ell} &= -\epsilon^{Ad}_{\ell\ell} = \frac{g_{Z'}^2}{2\sqrt{2} G_F m_{Z'}^2},
&
&(\ell=e,\mu,\tau).
\end{align}
The strongest bound from existing neutrino experiments stems from the NuTeV experiment, leading to the constraint $e^{dR}_{\mu\mu} > -0.01$ on $e^{dR}_{\mu\mu} = -\frac{2g_{Z'}^2}{3\sqrt{2} G_F m_{Z'}^2}$, at 95\% C.L. (see Tab.~\ref{tab:limits1}).  Generation-dependent charges can also be considered as in Ref.~\cite{Abbaslu:2024hep}.

As mentioned in the introduction, axial-vector NSIs cannot be probed through neutrino oscillations or elastic scattering, but inelastic neutrino-nucleus scattering is sensitive to these interactions. Therefore we apply the bounds from Tab.~\ref{tab:limits2} derived from projections for the DUNE ND. From Fig.~2 in Ref.~\cite{Abbaslu:2023vqk} one can see that bounds for $\epsilon^{Au}_{\ell\ell} = -\epsilon^{Ad}_{\ell\ell}$ are at least as strong as those for scenarios with only $\epsilon^{Au}_{\ell\ell}$ or $\epsilon^{Ad}_{\ell\ell}$, which implies that one can conservatively use the stronger bound of the two from the table.

The derivation of current and future collider bounds on this model proceeds in a similar fashion as in section~\ref{sec:ZBL}, using ATLAS and CMS studies for di-lepton resonances, both for existing LHC data~\cite{ATLAS:2019erb,ATLAS:2019erb_repo,CMS:2021ctt,CMS:2021ctt_repo} and HL-LHC projections~\cite{ATLAS:2018tvr,CMS:2022gho}, and calculations of expected sensitivities from $e^+e^- \to f\bar{f}$ and $\mu^+\mu^- \to f\bar{f}$ at future lepton colliders.

\begin{figure}[t]
    \centering
    \includegraphics[width=0.75\linewidth]{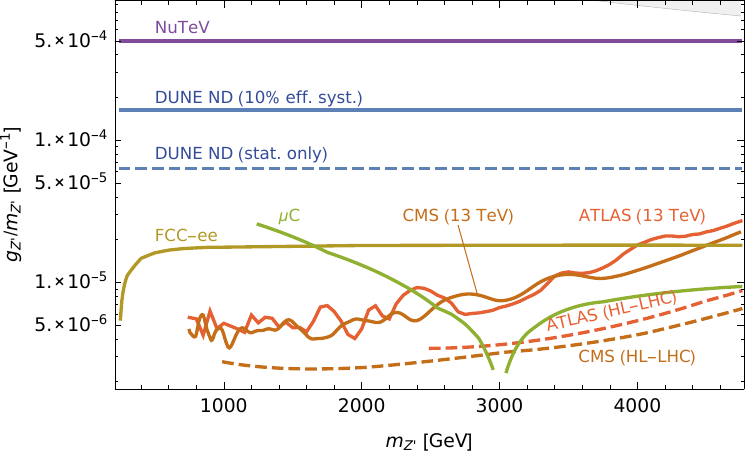}
    \caption{Comparison of current and future bounds (95\% C.L.) from neutrino experiments and colliders on $g_{Z'}/m_{Z'}$ for the $Z'$ model with axial charges.}
    \label{fig:ZpAx}
\end{figure}

The results are shown in Fig.~\ref{fig:ZpAx}. The region where the coupling $g_{Z'}$ becomes non-perturbative is the gray area shown in the plot. Compared to the U(1)$_{B-L}$ model, the projection for the DUNE ND reaches lower values of $g_{Z'}/m_{Z'}$, although the sensitivity strongly depends on assumptions about the systematics, as indicated by the two blue lines in the plot (the dashed lines corresponds to negligible systematics, whereas the solid assumes a 10\% systematic uncertainty on the detection efficiency). Nevertheless, the expected DUNE reach is still not competitive with existing bounds from LHC.


\subsection{Scalar leptoquarks}
\label{sec:leptoquark}

There are a number of minimal leptoquark models that can produce NSIs~\cite{Cherchiglia:2023aqp}, but their phenomenology is roughly similar in many ways. Therefore, for concreteness, this section considers a scalar leptoquark that is an electroweak doublet with hypercharge 1/6,
\begin{equation*}
    \Omega(3,2,\tfrac{1}{6}) = \begin{pmatrix} \omega^{2/3} \\ \omega^{-1/3} \end{pmatrix},
\end{equation*}
where the superscripts indicate the electric charge of the doublet components. [The $\Omega$ field is denoted $\Pi_1$ in Ref.~\cite{Cherchiglia:2023aqp}.] The phenomenology of this model in neutrino experiments and colliders has been studied in Ref.~\cite{Babu:2019mfe}. Here we extend this analysis by including additional constraints from neutrino scattering at the DUNE ND, from LHC Drell-Yan production, and from future lepton colliders.

The relevant part of the interaction Lagrangian can be written as
\begin{align}
    \cL^{\rm int}_{\rm lq,\Omega} &= - \sum_{\alpha=e,\mu,\tau} \lambda_{\alpha d} \,\overline{L}_\alpha \epsilon \Omega^* d + \text{h.c.},
\end{align}
where $\lambda_{\alpha d}$ is the leptoquark coupling for the lepton flavor $\alpha$ and $\epsilon$ is the antisymmetric tensor in SU(2) space.
To avoid constraints from lepton flavor physics, we only consider scenarios with one of the $\lambda_{\alpha d}$ couplings being non-zero. Specifically, we will examine the following scenarios:
\begin{itemize}
    \item Muon-philic leptoquark: $\lambda_{\mu d} \neq 0$, $\lambda_{e d} = \lambda_{\tau d} = 0$;
    \item Tau-philic leptoquark: $\lambda_{\tau d} \neq 0$, $\lambda_{e d} = \lambda_{\mu d} = 0$.
\end{itemize}
The NSI parameters are given by
\begin{align}
    \epsilon^{d}_{\ell\ell} = \epsilon^{Ad}_{\ell\ell} &= \frac{|\lambda_{\ell d}|^2}{4\sqrt{2}G_F m_\omega^2}, &
    \epsilon^{u}_{\ell\ell} = \epsilon^{Au}_{\ell\ell} &= 0,
\end{align}
where $m_\omega$ is the mass of $\omega^{2/3}$ and $\omega^{-1/3}$, which is degenerate at tree-level.

\medskip
For the muon-philic case, neutrino scattering places strong bounds on the parameter space, owing to the fact that accelerator neutrino beams are dominated by muon neutrinos. For existing data, the NuTeV bound $\epsilon^{dR}_{\mu\mu}<0.018$ (see Tab.~\ref{tab:limits1}) is most relevant. These can be superseded by future DUNE ND data, where the projected bounds on axial-vector NSIs from Tab.~\ref{tab:limits2} turn out to be the most sensitive.

The LHC constraints arise from three types of processes: First, leptoquarks can be produced in pairs. ATLAS~\cite{ATLAS:2020dsk} found a bound $m_\omega > 1740$~GeV for muon-philic leptoquarks produced in this mode (see Fig.~8 top-right there). One can derive a rough estimate of the HL-LHC reach by applying a luminosity scaling factor $\sqrt{\frac{139 \text{ fb}^{-1}}{3000 \text{ fb}^{-1}}} \approx 0.22$ on the expected cross-section limit. With this, we find a projected HL-LHC bound of $m_\omega > 1930$~GeV. 

Second, leptoquarks can also be produced singly, leading to the same $\mu\mu jj$ final state as pair production. Ref.~\cite{Bhaskar:2023ftn} presents a recast of the ATLAS result, combining both single and pair production modes. It also includes non-QCD production channels for the pair production mode. The limits for the $\Omega$ leptoquark can be taken from Fig.~4(k) of Ref.~\cite{Bhaskar:2023ftn}. [The $\Omega$ field is denoted $\tilde{R}_2$ there.] Since the dependence on the coupling and mass are quite different for the pair and single production modes, it is difficult to provide an estimated HL-LHC reach without a detailed breakdown of the analysis.

Third, leptoquarks can modify the rate for the Drell-Yan process, $pp\to\mu^+\mu^-$, through their $t$-channel contribution. This effect has also been studied in Ref.~\cite{Bhaskar:2023ftn} and the bounds from current data can be extracted from Fig.~4(k) there.\footnote{The Drell-Yan constraints were derived from a CMS study~\cite{CMS:2021ctt}, but there is also a similar ATLAS study~\cite{ATLAS:2020yat}. Comparable limits for $m_\omega \sim \text{few TeV}$ can also be obtained from the recasting study in Ref.~\cite{Allwicher:2022gkm}.} An estimate for the HL-LHC reach can be obtained by assuming that the leptoquark contribution to the Drell-Yan cross-section $\sigma_{\rm DY}$ is dominated by the interference ${\cal M}_{\rm SM}{\cal M}^*_{\rm LQ}$, where ${\cal M}_{\rm SM}$ and ${\cal M}_{\rm LQ}$ are the SM and leptoquark contributions to the matrix element, respectively. Thus $\sigma_{\rm DY}$ scales quadratically with $\lambda_{\mu d}$. Using naive luminosity scaling, the limit on $\sigma_{\rm DY}$ is improved by $\sqrt{\frac{140 \text{ fb}^{-1}}{3000 \text{ fb}^{-1}}}$, and thus the limit on $|\lambda_{\mu d}|$ improves by $\bigl(\frac{140 \text{ fb}^{-1}}{3000 \text{ fb}^{-1}}\bigr)^{1/4} \approx 0.47$.

At future $e^+e^-$ or $\mu^+\mu^-$ colliders, leptoquarks can be pair produced as long as $m_\omega < \sqrt{s}/2$, but this parameter region is already excluded by LHC data for the center-of-mass energies considered in this work. In addition, muon-philic leptoquarks can contribute to jet pair production at a muon collider via their $t$-channel exchange, which provides sensitivity to larger values of $m_\omega$. This contribution has been evaluated as described in section~\ref{sec:coll}.

\begin{figure}[t]
    \centering
    \includegraphics[width=0.75\linewidth]{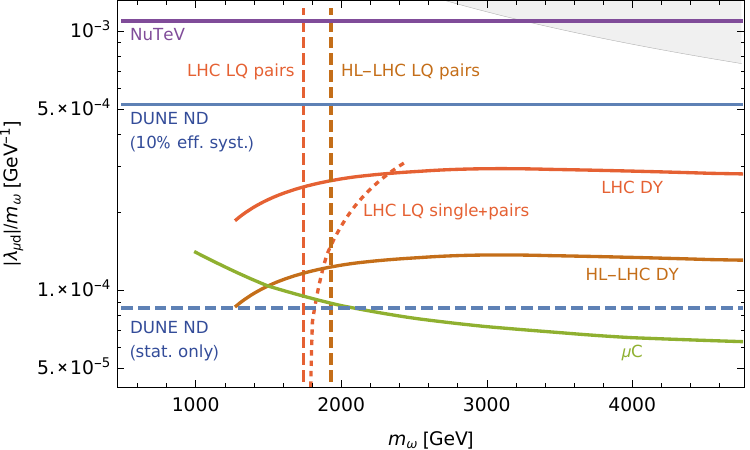}
    \caption{Bounds on $|\lambda_{\mu d}|/m_\omega$ for the scalar leptoquark $\Omega$ with $\lambda_{\mu d} \neq 0$, $\lambda_{ed} = \lambda_{\tau d}=0$ from current and future neutrino and collider experiments (95\% C.L.). The shaded region displays the perturbativity bound.}
    \label{fig:LQmu}
\end{figure}

Figure~\ref{fig:LQmu} shows the comparison of different current and projected bounds for the muon-philic $\Omega$ leptoquark. As can be seen from the plot, the projections for neutrino scattering at the DUNE ND with only statistical errors are competitive with the estimated reach of the HL-LHC and a muon collider, but the parameter space covered by the DUNE ND may be significantly degraded due to systematic uncertainties. Thus, a more careful analysis of experimental systematics would be crucial to better understand the capabilities of DUNE to probe leptoquark-mediated NSIs.

\medskip
For the tau-philic case, the current limit on $\epsilon^d_{\tau\tau}$ from neutrino oscillation and CE$\nu$NS experiments is given in Tab.~\ref{tab:limits1}~(left). Since $\epsilon^d_{\tau\tau} >0$ in the scalar leptoquark model, the upper bound $\epsilon^d_{\tau\tau} < 0.051$ applies. This limit is not expected to be improved by future neutrino oscillation data~\cite{Liao:2016orc,Chatterjee:2021wac,Du:2021rdg,Cherchiglia:2023aqp} or DUNE far-detector scattering data~\cite{Abbaslu:2023vqk}.  The projected KM3NeT/ORCA bound \cite{HernandezRey:2021qac}, $|\epsilon_{\tau\tau}^\oplus| < 0.0134$ at 95\% C.L., leads to a constraint $|\epsilon_{\tau\tau}^d| < 0.0043$.

From the LHC, there are bounds from leptoquark pair production, for which we use the limit $m_\omega > 990$~GeV from Fig.~3(a) of Ref.~\cite{CMS:2018qqq}. As before, a rough estimate of the HL-LHC pair production bound can be obtained by using a luminosity scaling factor $\sqrt{\frac{35.9 \text{ fb}^{-1}}{3000 \text{ fb}^{-1}}} \approx 0.11$ on the cross-section limit.

In addition, similar to above, a tau-philic leptoquark would modify the rate for Drell-Yan production, $pp\to\tau^+\tau^-$. Ref.~\cite{Allwicher:2022gkm} gives a bound for $m_\omega = 2$~TeV in Fig.~4.7 there (for $[\tilde{y}^L_2]_{13} \neq 0$ in the notation of that reference).  To derive limits for other leptoquark masses, the cross-section for $pp\to\tau\tau$ has been computed in {\sc CalcHEP} for different values of $m_\omega$ using the cut $m_T^{\rm tot} > 200$~GeV (see Tab.~4.1 in Ref.~\cite{Allwicher:2022gkm}). For the projected HL-LHC bounds, the luminosity scaling as described above for the muon-philic case has been used.

Both a future $e^+e^-$ collider like FCC-ee or a 3-TeV muon collider do not provide additional coverage of the tau-philic leptoquark beyond HL-LHC.

\begin{figure}[t]
    \centering    \includegraphics[width=0.75\linewidth]{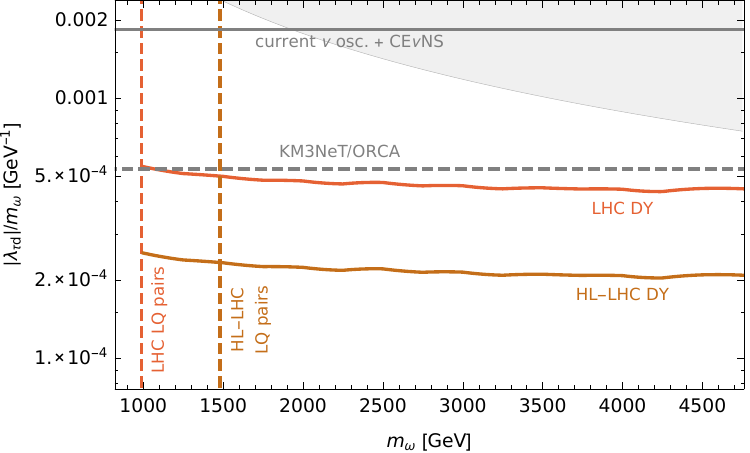}
    \caption{Bounds on $|\lambda_{\tau d}|/m_\omega$ for the scalar leptoquark $\Omega$ with $\lambda_{\tau d} \neq 0$, $\lambda_{ed} = \lambda_{\mu d}=0$ from current and future neutrino and collider experiments (95\% C.L.). The shaded region displays the perturbativity bound.}
    \label{fig:LQtau}
\end{figure}

As shown in Fig.~\ref{fig:LQtau}, the current LHC bounds on the tau-philic $\Omega$ leptoquark already surpass the limits from NSIs at neutrino experiments by a significant margin. While new ideas for future neutrino experiments may improve the reach for NSI parameters, it is unlikely that for this model they will probe new regions of parameter space beyond the (HL-)LHC coverage.

While we study a scalar leptoquark that only couples to down quarks we expect that any scalar or vector leptoquark that couples to combinations of first generation quarks to have qualitatively similar results.  Couplings to second and third generation quarks would fully remove constraints from neutrino experiments while the pair-production limits from the LHC and HL-LHC would only be minimally affected.


\subsection{Heavy neutral leptons}
\label{sec:HNL}

Heavy neutral leptons (HNLs) with vector-like mass terms contribute to NSIs through their mixing with SM neutrinos. In general, HNLs can occur in different SU(2) representations, but here we focus on SU(2) singlets, since the collider constraints on larger representations are generally stronger.

The interaction and mass terms of such a singlet HNL are given by
\begin{align}
    \cL^{\rm int,m}_{\rm HNL} &= -\sum_{\alpha=e,\mu,\tau} y_\alpha \overline{L}_\alpha H N_R - m_N \overline{N}_L N_R + \text{h.c.},
\end{align}
where $N_{L,R}$ are the left-/right-handed part of the pseudo-Dirac HNL.
Below, we will consider different scenarios for the flavor pattern of the Yukawa coupling $y_\alpha$. In the limit of $v \ll m_N$, where $v$ is the Higgs vev and $m_N$ is the HNL mass, the $\nu_\alpha$--$N$ mixing angle is
\begin{align}
    \theta_\alpha \approx \frac{y_\alpha v}{\sqrt{2}\,m_N}.
\end{align}
This mixing effect contributes to low-energy NSIs via the operator $\cO_{HL}^{(1)}$ in Eq.~\eqref{eq:OHl}, in contrast to the previous models that generate four-fermion SMEFT operators.

It is important to note that the Fermi constant is also modified in this model, $G_F = G_F^{\rm SM}(1-\theta_e-\theta_\mu)$, which has an impact on the NSIs in Eqs.~\eqref{eq:NC} and \eqref{eq:CC} on account of their normalization with respect to $G_F$. One then finds
\begin{align}
    \epsilon^{u}_{\ell\ell} &= \Bigl(\frac{1}{4}-\frac{2}{3}s_W^2\Bigr) (\theta_e^2+\theta_\mu^2-2\theta_\ell^2), &
    \epsilon^{d}_{\ell\ell} &= \Bigl(-\frac{1}{4}+\frac{1}{3}s_W^2\Bigr) (\theta_e^2+\theta_\mu^2-2\theta_\ell^2), \\
    \epsilon^{Au}_{\ell\ell} &= -\frac{1}{4}(\theta_e^2+\theta_\mu^2-2\theta_\ell^2), &
    \epsilon^{Ad}_{\ell\ell} &= \frac{1}{4}(\theta_e^2+\theta_\mu^2-2\theta_\ell^2).
\end{align}
There is also a modification of charged-current NSIs, 
\begin{align}
    \bar{\epsilon}_{\ell\ell}^{qL} = \theta_e^2 + \theta_\mu^2 - \theta_\ell^2. \label{eq:CCmod}
\end{align}
Similarly to the discussion in the previous section, neutrino scattering can set strong bounds on $\epsilon^q_{\mu\mu}$ and $\epsilon^{Aq}_{\mu\mu}$.  From available NuTeV data, the bound $|\epsilon_{\mu\mu}^{dL}| < 0.004$ (see Tab.~\ref{tab:limits1}) is the most constraining, leading to
\begin{align}
    |\theta_e^2-\theta_\mu^2| < 0.019 \qquad \qquad \text{(NuTeV)}.
\end{align}
For DUNE ND projections, we again find the bounds on axial-vector NSIs from Tab.~\ref{tab:limits2} to be the most sensitive.

In addition, several future experiments can probe the modified charged-current interaction in Eq.~\eqref{eq:CCmod} through pion and beta decay. The projected limits are~\cite{Du:2021rdg}
\begin{align}
    &|\bar\epsilon^{qL}_{ee}| < 2\times 10^{-3} \,\qquad\qquad \text{(JUNO/TAO)}, \label{eq:beta} \\
    &|\bar\epsilon^{qL}_{\mu\mu}| < 2\times 10^{-2} \qquad\qquad \text{(T2HK/DUNE)}. \label{eq:pion}
\end{align}
If $\theta_e=\theta_\mu=0$, neutrino scattering constraints from NuTeV and the DUNE ND and beta decay bounds do not apply.  There is a bound from oscillation data and CE$\nu$NS, see Tab.~\ref{tab:limits1}~(left), leading to $\theta_\tau^2 < 0.15$ with present data, which is not expected to be improved upon by future data from DUNE, T2HK, JUNO, and TAO~\cite{Liao:2016orc,Chatterjee:2021wac,Du:2021rdg,Cherchiglia:2023aqp,Abbaslu:2023vqk}. A stronger projected bound is obtained from KM3NeT/ORCA~\cite{HernandezRey:2021qac}, which leads to $\theta_\tau^2 < 0.022$.

At colliders, HNLs can be produced in association with a SM lepton through weak interactions. At the LHC, the main production process is charged-current Drell-Yan, $pp \to W^{\pm *} \to N \ell^\pm$, followed by the decay $N \to \ell\ell'\nu_{\ell''}$, where different flavor combinations of the final-state leptons are possible. We use bounds obtained by CMS with 138~fb$^{-1}$ at $\sqrt{s}=13$~TeV~\cite{CMS:2024xdq}. To obtain projections for the HL-LHC one can again use a simple luminosity scaling factor $\bigl(\frac{138 \text{ fb}^{-1}}{3000 \text{ fb}^{-1}}\bigr)^{1/4}$ on the mixing angle limit. There are also HL-LHC projections from Ref.~\cite{Antusch:2016ejd}, which lead to similar bounds.

At a future multi-TeV muon collider, HNLs can be produced through $\mu^+\mu^- \to N\nu$, followed by the decay $N \to W^\pm\ell^\mp$, where hadronically decaying W bosons offer the best sensitivity. This has been studied in Ref.~\cite{Li:2023tbx}, and projected 95\% C.L. bounds have been derived there. We do not consider direct search projections for future $e^+e^-$ Higgs factories due to the limited mass reach.

Nevertheless, electroweak precision measurements can put significant bounds on HNLs by constraining the mixing angle of the SM-like neutrinos. Here, the most relevant quantities are the total $Z$ width, which would be reduced by $\nu_\alpha$--$N$ mixing, and the Fermi constant, which was already mentioned above. We use bounds derived from current data from Ref.~\cite{Blennow:2023mqx}, and projections for FCC-ee from Ref.~\cite{Antusch:2016ejd}.

\begin{figure}[p]
    \centering
    \begin{tabular}{c@{\hspace{-1em}}r}
    a) \\[-1em] & \includegraphics[height=2.4in]{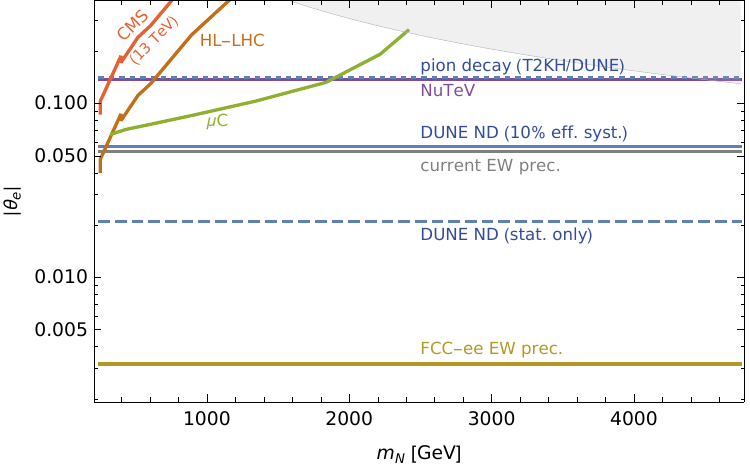}\\[-1ex]
    b) \\[-1em] & \includegraphics[height=2.4in]{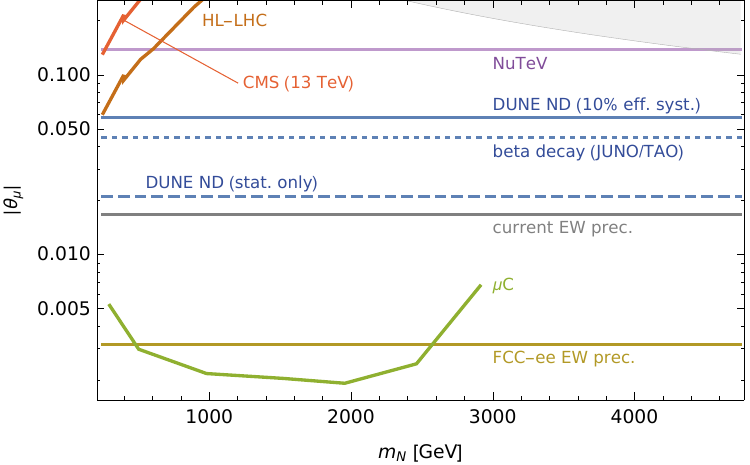}\\[-1ex]
    c) \\[-1em] & \includegraphics[height=2.4in]{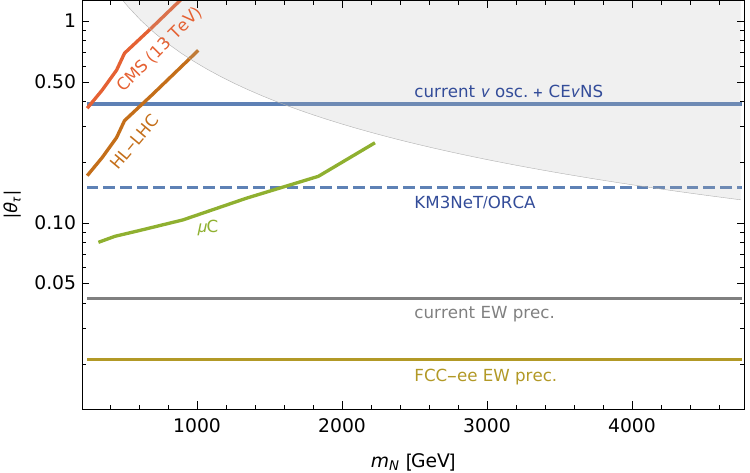}
    \end{tabular}
    \vspace{-1ex}
    \caption{Bounds on the mixing angle between active neutrinos and a HNL $N$ from current and future neutrino and collider experiments (95\% C.L.) for different flavor patterns: $N$ mixing only with $\nu_e$ (a), only with $\nu_\mu$ (b), and only with $\nu_\tau$ (c).  
    The shaded regions show the perturbativity bounds.}
    \label{fig:HNL}
\end{figure}

The comparison of sensitivities in HNL parameter space of current and future neutrino and collider experiments are shown in Fig.~\ref{fig:HNL}, for three scenarios: (a) only $\nu_e$--$N$ mixing; (b) only $\nu_\mu$--$N$ mixing; and (c) only $\nu_\tau$--$N$ mixing. The (HL-)LHC bounds in the plots are for the case that the sterile neutrino $N$ is a Dirac particle, but the exclusions for a Majorana $N$ are roughly similar~\cite{CMS:2024xdq}, and thus they are not shown separately. The (HL)-LHC constraints for these HNL models are rather weak, and the strongest collider limits are generally obtained from electroweak precision data.  Only in scenario (b), \emph{i.e.}\ mixing with the muon neutrino, do direct HNL searches at a muon collider exceed the reach of projected electroweak precision data from an $e^+e^-$ collider like FCC-ee for some range of $m_N$ values. 

Bounds on the HNL mixing angles from NSIs at neutrino experiments generally exceed the current and projected reach of the LHC. However, for the most part, electroweak precision data can constrain these models more tightly than results from neutrino experiments. An exception is the case (a), where the statistical precison for DUNE ND scattering reaches lower values of $|\theta_e|$
than the current electroweak precision tests and muon collider searches. This conclusion, however, strongly depends on the systematic uncertainties of the DUNE ND measurements, as it turns out that a 10\% uncertainty on the efficiency would significantly deteriorate the reach for small mixing angles. Thus, it is important to have a more careful analysis of systematic uncertainties for neutrino scattering at the DUNE ND. Nevertheless, we note that electroweak precision tests at a future FCC-ee would yield stronger limits then DUNE or other neutrino experiments in all three scenarios.

For scenario (c), \emph{i.e.}\ $\nu_\tau$--$N$ mixing, the only relevant bounds from neutrino experiments are $\theta_\tau^2<0.15$ from a combination of existing oscillation data and CE$\nu$NS, as discussed above, and the projected limit from KM3NeT/ORCA, $\theta_\tau^2 < 0.022$.

Finally, we also want to briefly comment on a scenario where $N$ mixes democratically with all SM neutrinos, $\theta_e=\theta_\mu=\theta_\tau \neq 0$. In this case, all NC NSI parameters vanish, $\epsilon^q_{\ell\ell} = \epsilon^{Aq}_{\ell\ell} = 0$. However, for the CC NSI parameters one finds $\bar\epsilon^{qL}_{\ell\ell} = \theta$, so that the bounds from pion and beta decay in Eqs.~\eqref{eq:beta} and~\eqref{eq:pion} still apply, even though they are weaker than bounds from electroweak precision tests and muon collider searches.


\section{NSIs from dimension-8 operators}
\label{sec:dim8}

In this section, we wish to add more detail to the analysis of the possibility that neutrino NSIs are generated by dimension-8 SMEFT operators of the form in Eq.~\eqref{eq:OHlq}, in order to avoid strong constraints from observables with charged leptons.
A scenario where the dominant contribution to NSIs stems from dimension-8 operators requires that the Wilson coefficients of the dimension-6 operators Eqs.~\eqref{eq:Olq}--\eqref{eq:OHl3} are zero or strongly suppressed. The work in Ref.~\cite{Gavela:2008ra} proved that there is no model with a single tree-level mediator that generates the dimension-8 operator Eq.~\eqref{eq:OHlq} without also producing dimension-6 contributions to NSIs.

\begin{figure}[t]
    \centering
    \begin{tabular}{|c|c|}
    \hline
    Contribution to dimension-8 operator &
    Contribution to dimension-6 operator \\
    \hline
    \includegraphics[width=0.27\linewidth]{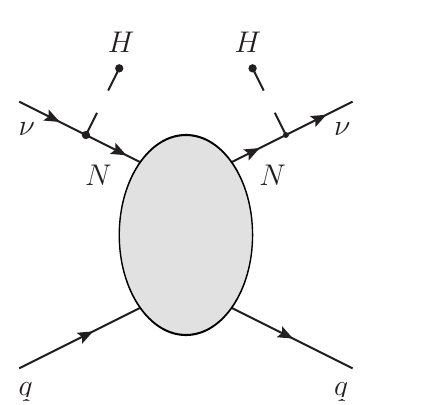}
    &
    \includegraphics[width=0.27\linewidth,height=0.27\linewidth]{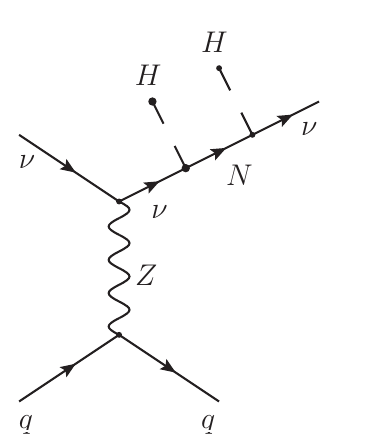} \\
    \hline
    \end{tabular}
    \caption{Diagrams of a dimension-8 operator (left) and a dimension-6 operator (right) that is generated by a heavy neutral lepton $N$.  The $\nu$--$N$ mixing generates the dimension-6 operator ${\cal O}^{(1)}_{HL}$ at tree-level.}
    \label{fig:dim-diagram-HNL}
\end{figure}

\begin{figure}[t]
    \centering
    \begin{tabular}{|c|c|}
    \hline
    Contribution to dimension-8 operator &
    Contribution to dimension-6 operator \\
    \hline
    \includegraphics[width=0.27\linewidth]{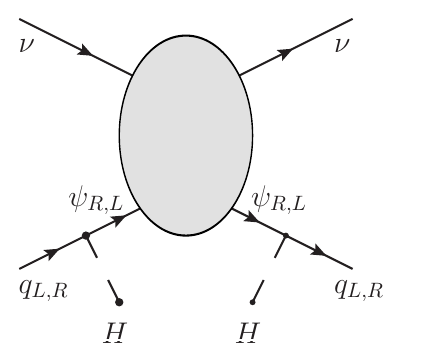}
    &
    \includegraphics[width=0.27\linewidth]{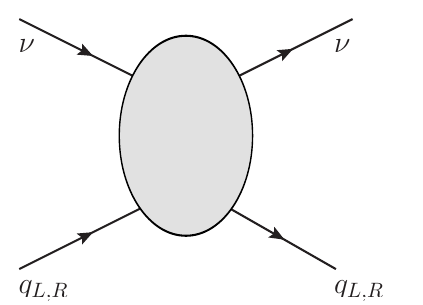} \\
    \hline
    \end{tabular}
    \caption{Diagrams of a dimension-8 operator (left) and a dimension-6 operator (right) that is generated by the new fermions $\psi_{R,L}$, which have the same quantum numbers as $q_{R,L}$.  Since $\psi_{R,L}$ and $q_{R,L}$ have the same quantum numbers, whatever couples to $\psi_{R,L}$ in the dimension-8 operator will also couple to $q_{R,L}$, leading to the dimension-6 operators ${\cal O}^{(1,3)}_{Lq}$.}
    \label{fig:dim-diagram-psi}
\end{figure}

\begin{figure}[t]
    \centering
    \begin{tabular}{|c|c|}
    \hline
    Contribution to dim.-8 operator &
    Contributions to dimension-6 operator \\
    \hline
    \includegraphics[width=0.3\linewidth]{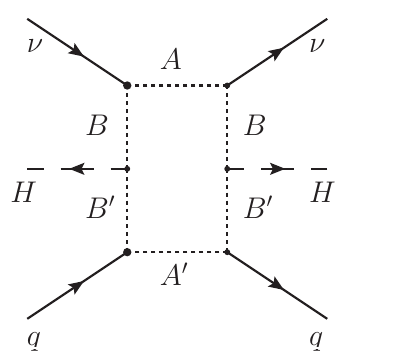}
    &
    \includegraphics[width=0.27\linewidth]{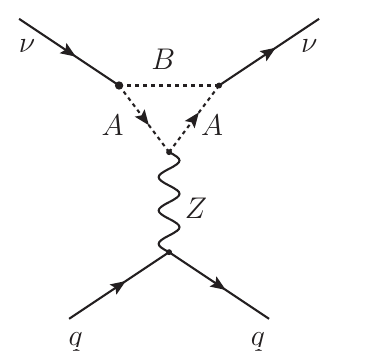} 
    \includegraphics[width=0.27\linewidth]{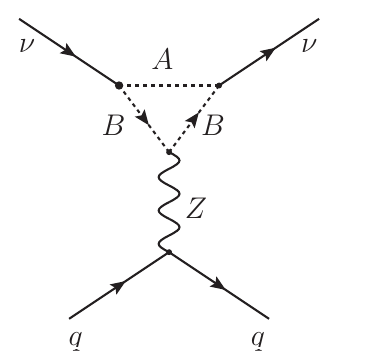} \\
    \hline
    \end{tabular}
    \caption{Diagrams of a dimension-8 operator (left) and dimension-6 operators (right) that are generated by the new bosons and/or fermions $A$ and $B$.  In the simplest case, one is an electroweak doublet and the other an electroweak singlet; however in general as long as they couple to an electroweak doublet, they have can have any representations.  The first diagram on the right is generated when $A$ is an electroweak doublet (or higher representation) and the second diagram on the right is generated when $B$ is an electroweak doublet (or higher representation).  These correspond to the operators ${\cal O}^{(1,3)}_{HL}$.}
    \label{fig:dim-diagram-scalar}
\end{figure}

According to our analysis, a similar conclusion holds for simple models with loop-induced NSI operators. Here ``simple'' refers to the assumption that there are no cancellations between loop diagrams with different particle content. The basic argument is that any loop diagram contributing to Eq.~\eqref{eq:OHlq} needs to include at least one BSM field that is charged under the weak SU(2). This field (or these fields) would therefore couple to the $Z$ boson and induce one or both of the operators in Eqs.~\eqref{eq:OHl} and~\eqref{eq:OHl3}.  This is illustrated in Figs.~\ref{fig:dim-diagram-HNL}--\ref{fig:dim-diagram-scalar}.

Consequently, the only possibility to generate dimension-8 but no dimension-6 contributions to NSIs is through fine-tuned cancellations between two or more diagrams with different mediator fields. Such a possibility might not be very elegant, but it should not be discounted outright, since fine-tuning at the level of several percent is sufficient to suppress the dimension-6 contributions to a subdominant level. A suitable toy model has been presented in Ref.~\cite{Gavela:2008ra} for leptonic NSIs, and in the following we consider the analogue of this model for neutrino-quark NSIs.

The model contains a scalar leptoquark $\Omega(3,2,\frac{1}{6})$ and a vector leptoquark $V(3,2,-\frac{5}{6})$, where the numbers in parentheses indicate the representations under color SU(3), weak SU(2), and the hypercharge, respectively. The relevant part of the interaction Lagrangian is given by
\begin{align}
    \cL^{\rm int}_{\rm dim8} = -& \sum_{\alpha=e,\mu,\tau} \big[ \lambda_{\alpha d} \,\overline{L}_\alpha \epsilon \Omega^* d +g_{\alpha d} \,\overline{L}_\alpha \gamma^\mu V_\mu d^c + \text{h.c.} \bigr] \notag \\
    -& \; \kappa_{\rm s}(H^\dagger H)(\Omega^\dagger \Omega) - \kappa_{\rm v}(H^\dagger H)(V^\dagger_\mu V^\mu), \label{eq:LQSV}
\end{align}
where $H$ is the SM Higgs doublet field.

The contribution from dimension-6 SMEFT operators to the NSI parameters is given by
\begin{align}
    \epsilon^{d}_{\ell\ell} = \epsilon^{Ad}_{\ell\ell} &= \frac{1}{2\sqrt{2}G_F}
    \biggl( \frac{|\lambda_{\ell d}|^2}{2m_\Omega^2} - \frac{|g_{\ell d}|^2}{M_V^2} \biggr),
\end{align}
which vanishes if
\begin{align}
    M_\Omega &= M_V \equiv M, \qquad
    |\lambda_{\ell d}|^2 = 2|g_{\ell d}|^2. \label{eq:dim6canc}
\end{align}
Assuming that Eq.~\eqref{eq:dim6canc} holds, the contribution from the dimension-8 operator Eq.~\eqref{eq:OHlq} is~\cite{Gavela:2008ra}
\begin{align}
    \epsilon^{d}_{\ell\ell} = \epsilon^{Ad}_{\ell\ell} &=
    \frac{|g_{\ell d}^2|(\kappa_{\rm s}+\kappa_{\rm v})}{4G_F^2 M^4}
\end{align}
Therefore, compared to the scalar leptoquark model in section~\ref{sec:leptoquark}, the bounds from neutrino experiments on $|\lambda|/M$ are only weakened by a factor $M/(v\sqrt{\kappa_{\rm s}+\kappa_{\rm v}})$. 

On the other hand, the bounds from leptoquark pair production at the LHC in section~\ref{sec:leptoquark} would approximately still apply, since they are governed by gauge interactions. However, the limits from Drell-Yan production are modified. To evaluate this modification, the scalar+vector leptoquark toy model has been implemented in {\tt CalcHEP} and cross-section results for $pp \to \ell^+\ell^-$ have been produced using the cuts and binning in Ref.~\cite{CMS:2021ctt}, and the condition in Eq.~\eqref{eq:dim6canc}. It turns out that for $M \lesssim 2$~TeV, the contribution to Drell-Yan from the toy model is comparable to the scalar leptoquark model in section~\ref{sec:leptoquark}. With increasing $M$, the cancellation between the scalar and vector leptoquark contributions becomes more significant, leading to weakened LHC bounds. For example, for $M=5$~TeV, the LHC Drell-Yan limits on $|\lambda|/M$ are degraded by a factor $\approx 1.5$ compared to section~\ref{sec:leptoquark}.

Nevertheless, this constitutes only a relatively minor relaxation of the LHC Drell-Yan bounds. Given the fact that the observable NSI effects at neutrino experiments are also reduced in the toy model, since they are produced only by a dimension-8 operator, we conclude that the viable parameter space in this model is similarly or even more stringently constrained by collider data than the scalar leptoquark model in section~\ref{sec:leptoquark}.


\section{Summary}
\label{sec:summ}

Neutrino non-standard interactions (NSIs) can be mediated by massive beyond-the-Standard-Model (BSM) particles. While neutrino experiments can probe NSIs in the form of low-energy effective four-fermion interactions, high-energy colliders may directly produce the heavy BSM mediators. This paper compares the current bounds and projected future reach of neutrino experiments and colliders for a broad range of simple representative models that generate neutrino-quark NSIs at tree-level through dimension-6 effective operators. 

For the neutrino experiments, a combination of currently available neutrino oscillation and scattering data has been taken into account, and future projections for a number of planned facilities have been considered, in particular the DUNE experiment. In case of NSIs involving muon neutrinos, neutrino scattering data, \emph{e.g.}\ from NuTeV and the DUNE near-detector (ND), lead to stronger bounds than neutrino oscillation results.
For the collider experiments, bounds have been derived from LHC and LEP data, and we have estimated the expected improvement from the full HL-LHC dataset, an $e^+e^-$ Higgs factory (using FCC-ee as a concrete example), and a multi-TeV muon collider.

Three main classes of simple models have been considered: spin-0 mediators (leptoquarks), spin-1/2 mediators (heavy neutral leptons), and spin-1 mediators (gauge bosons), which all can produce sizable NSIs even for TeV-scale masses. To avoid constraints from flavor observables, it has been assumed that the couplings of these mediators conserve both quark and lepton flavor.

It turns out that for most of these simple models, existing LHC and LEP data already exclude the parameter space where current and future neutrino experiments have sensitivity. Future data from the HL-LHC or a muon collider would further strengthen the bounds. There are some notable exceptions, however: 
\begin{itemize}
    \item[a)] Muon-philic leptoquarks, where DUNE ND scattering data, with only statistical uncertainties, can place stronger bounds than the HL-LHC for leptoquark masses $m\gtrsim 2$~TeV (see Fig.~\ref{fig:LQmu}). The DUNE ND sensitivity could also be competitive with the reach of a 3-TeV muon collider.
    \item[b)] Heavy neutral leptons (HNLs) that mix with electron-neutrinos, where LHC bounds are weak, and the statistical sensitivity of the DUNE ND would exceed the bounds from current (LEP) electroweak precision data (Fig.~\ref{fig:HNL}), even though not the projected reach of the FCC-ee electroweak precision program.
\end{itemize}
Note, however, that the projected sensitivity of DUNE ND scattering data for probing NSIs could be strongly degraded due to systematic uncertainties. Therefore, it is very important to more rigorously evaluate the expected size of systematic errors, in order to assess the potential of DUNE ND data to cover parameter regions beyond the reach of colliders (in particular for the leptoquark and HNL models).

Finally, we also briefly discussed NSIs that are generated by dimension-8 electroweak effective operators. 
Such scenarios have been proposed to evade or weaken bounds from observables involving charged leptons. However, as was previously pointed out, one needs fine-tuned cancellations between two (or more) mediators in any concrete model realization to suppress the relevant dimension-6 operators and obtain a dominant contribution from a dimension-8 operator. We have considered the collider constraints for an example of such a fine-tuned model with two leptoquark mediators and found that the cancellation between the two mediators is much less effective at LHC energies, and thus this model is still strongly constrained by LHC Drell-Yan data.


\section*{Acknowledgments}

The authors would like to thank Tao Han, Gregor Daberstiel and Dominik St\"ockinger for useful discussions and feedback.
This work has been supported in part by the U.S.~National Science Foundation Grant No.\ PHY-2412696 and by the US Department of Energy Grant No.\ DE-SC0007914.


\bibliographystyle{JHEP}{}
\bibliography{nsi}

\end{document}